\documentclass[reprint,superscriptaddress,showpacs,twocolumn,groupedaddress,amsmath,amssymb,aps,prl]{revtex4-1}
\usepackage{graphicx}  
\usepackage{dcolumn}   
\usepackage{bm}        
\usepackage{tabularx}
\usepackage{array}

\usepackage{float}
\usepackage[caption = false]{subfig}

\usepackage{color} 
\usepackage{ulem} 
\usepackage{hyperref}
\DeclareMathOperator{\sgn}{sgn}

\begin{document}

\title{Persistent ferromagnetism and topological phase transition at the interface of a superconductor and a topological insulator}

\author{Wei Qin}
\author{Zhenyu Zhang}
\affiliation{International Center for Quantum Design of Functional Materials (ICQD), Hefei National Laboratory for Physical Sciences at Microscale (HFNL), and Synergetic Innovation Center of Quantum Information and Quantum Physics, University of Science and Technology of China, Hefei, Anhui, 230026, China}

\date{\today}

\begin{abstract}
At the interface of an s-wave superconductor and a three-dimensional topological insulator, Majorana zero modes and Majorana helical states have been proposed to exist respectively around magnetic vortices and geometrical edges. Here we first show that a single magnetic impurity at such an interface splits each resonance state of a given spin channel outside the superconducting gap, and also induces two new symmetric impurity states inside the gap. Next we find that an increase in the superconducting gap suppresses both the oscillation magnitude and period of the RKKY interaction between two interface magnetic impurities mediated by BCS quasi-particles. Within a mean field approximation, the ferromagnetic Curie temperature is found to be essentially independent of the superconducting gap, an intriguing phenomenon due to a compensation effect between the short-range ferromagnetic and long-range anti-ferromagnetic interactions. The existence of persistent ferromagnetism at the interface allows realization of a novel topological phase transition from a non-chiral to a chiral superconducting state at sufficiently low temperatures, providing a new platform for topological quantum computation.
\end{abstract}

\maketitle
\textit{Introduction}.---Non-Abelian fermions have attracted much attention because of their potential applications in topological quantum computation (TQC) \cite{A. Kitaev,C. Nayak}. One common physical entity obeying non-Abelian braiding statistics is the zero-energy Majorana fermion \cite{G. Moore}, which is its own anti-particle described by $\gamma=\gamma^{\dagger}$. In condensed matter physics, a chiral topological superconductor (TSC) \cite{X. L. Qi1} is characterized by the existence of two types of Majorana fermions, chiral Majorana edge modes and a single Majorana zero mode surrounding a magnetic vortex, the latter can be manipulated for realization of TQC \cite{L. J. Buchholtz,M. Matsumoto,X. L. Qi2,D. A. Ivanove}. The simplest chiral TSC is a spinless $p_x+ip_y$ superconductor or superfluid \cite{N. Read}; however, it is difficult to quench the spin degrees of freedom in order to realize spinless superconductors.

Recently, Fu and Kane proposed that the proximity-induced superconductivity on the surface of a topological insulator (TI) deposited on a conventional s-wave superconductor possesses a $p_x+ip_y$ pairing feature \cite{L. Fu}.  The non-chiral nature of such a spinfull superconductor is characterized by the existence of Majorana helical edge states and a pair of Majorana zero modes surrounding a magnetic vortex. To convert such a TSC into a chiral one, time reversal symmetry (TRS) must be broken. Two schemes have been proposed to break TRS, both relying on the effect of a Zeeman field. The first consists of a superconductor-TI-magnet junction \cite{L. Fu}; in the second scheme, the TI can further be replaced by a traditional semiconducting thin film with strong Rashba spin-orbit coupling (SOC) \cite{J. D. Sau, J. Alicea}. These intriguing proposals have motivated extensive experimental efforts for the detection of Majorana fermions \cite{V. Mourik,A. Das,M. T. Deng}, but so far definitive proofs of their existence remain controversial. Here we note that both schemes face the inherent challenge that the proximity-induced Zeeman field decays rapidly through the TI or semiconductor thin film.

In this Letter, we introduce an alternative and conceptually new scheme to realize a chiral TSC within a simpler structure, achieved by doping magnetic impurities directly at a superconductor-TI interface. We first show that a single magnetic impurity at such an interface splits each resonance state of a given spin channel outside the superconducting gap, and also induces two new symmetric impurity states inside the gap. Next we find that an increase in the superconducting gap suppresses both the oscillation magnitude and period of the Ruderman-Kittel-Kasuya-Yosida (RKKY) interaction between two magnetic impurities mediated by BCS quasi-particles. The ferromagnetic Curie temperature is found to be essentially independent of the superconducting gap, due to a compensation effect between the short-range ferromagnetic and long-range anti-ferromagnetic interactions. The existence of persistent ferromagnetism at the interface provides a strong and uniform Zeeman field for the realization of a chiral TSC. In particular, by investigating the edge states and the corresponding first Chern number \cite{D. Xiao} , we reveal a topological phase transition from a non-chiral to chiral TSC at sufficiently low temperatures. These findings in principle provide a new and more appealing platform for TQC.

\textit{Theoretical model}.---The surface states of strong TIs are described by the time reversal invariant Hamiltonian $H_0=\sum_{k} \psi_k^{\dagger}(\nu_F \vec{\sigma} \cdot \vec{k}) \psi_k$. Here $\psi_k^{\dagger}=(c_{k\uparrow}^{\dagger},c_{k\downarrow}^{\dagger})$, $\vec{\sigma}=(\sigma_x,\sigma_y)$ are the Pauli spin matrices, $\mu$ is the chemical potential, and $\nu_F$ is Fermi velocity, given by 4.08 eV$\cdot$\AA~for Bi$_2$Se$_3$ \cite{H. J. Zhang} and 3.70 eV$\cdot$\AA~for Sb$_2$Te$_3$ \cite{Q. Liu}. By depositing an s-wave superconductor on the surface of a TI, the proximity-induced pairing Hamiltonian is given as $H_p=\sum_k(\Delta c_{k\uparrow}^{\dagger} c_{-k\downarrow}^{\dagger}+h.c.)$. Here $\Delta=\Delta_0 e^{i\phi}$ is the superconducting gap with phase $\phi$. The states at the superconductor-TI interface can then be described by \cite{L. Fu}
\begin{equation}
\label{Interface Hamiltonian}
\begin{aligned}
H_0&=\frac{1}{2} \sum_{k} \Psi_{k}^{\dagger} \mathcal{H}(\vec{k}) \Psi_{k}, \\
\mathcal{H}(\vec{k})&=(\nu_F\vec{\sigma} \cdot \vec{k}-\mu) \tau_z -\Delta_0(\tau_x \cos{\phi}-\tau_y \sin{\phi}),
\end{aligned}
\end{equation}
where $\Psi_{k}^{\dagger}=(c_{k\uparrow}^{\dagger},c_{k\downarrow}^{\dagger},c_{-k\downarrow},-c_{-k\uparrow})$ are 4-dimensional field operators in Nambu spinor basis. TRS and particle-hole symmetry are expressed as $\Theta=i \sigma_y K$ and $\Xi=\sigma_y \tau_y K$, which satisfy $[\Theta, \mathcal{H}]=0$ and $\{\Xi,\mathcal{H}\}=0$ at $\Gamma$ point of the Brillouin zone,  respectively, where $K$ is the complex conjugate operator. 

At the microscopic level, we treat the $s$-$d$ interaction between a magnetic impurity located at $\vec{R_i}$ and the electrons at the superconductor-TI interface to be isotropic, described by $H_{sd}^i=-J(\vec{\sigma} \cdot \vec{S}) \delta(\vec{r}-\vec{R_i})$, where $\vec{\sigma}=(\sigma_x,\sigma_y,\sigma_z)$ is the real electron spin, and $\vec{S}$ is the spin of the magnetic impurity. In Nambu notations, the interaction Hamiltonian can be rewritten as
\begin{equation}
\label{s-d interaction}
H_{sd}=-\frac{J}{2}\sum_{kk'}\Psi_{k}^{\dagger} (\vec{S} \cdot \vec{\sigma}) \tau_0 \Psi_{k'},
\end{equation}
where $J$ denotes the $s$-$d$ exchange coupling strength at the interface, estimated to be $0.1-0.5$ eV \cite{Q. Liu,J. S. Dyck,J. J. Zhu}. Hamiltonian (2) describes the interaction between the magnetic impurities and BCS quasi-particles, which, together with Hamiltonian (1) define our theoretical model and the starting point of this study.

\textit{Single magnetic impurity}.---We first study a single magnetic impurity at the superconductor-TI interface. The matrix form of the retarded Green's function of Eq.~(1) reads
\begin{equation}
G_0^{ret}(\vec{k},\omega)=\frac{1}{\omega-\mathcal{H}(\vec{k})+i\delta}.
\end{equation}
In order to study the effect of a single magnetic impurity, we investigate the local density of states (LDOS) using the $T$-matrix technique \cite{A. V. Balatsky}, which can be expressed using the Lippmann-Schwinger equation:
\begin{equation}
\hat{T}(\omega)=\hat{U}+\hat{U}G_0^{ret}(\omega,0)\hat{T}(\omega),
\end{equation}
where $G_0^{ret}(\omega,0)$ is the retarded Green's function in real space and $\hat{U}=-\frac{J}{2}(\vec{S}\cdot \vec{\sigma})\tau_0$ in our system. The algebra is simplest for $\mu=0$, where the retarded Green's function in real space is given by the Fourier transformation of Eq.~(3). For $|\omega|>\Delta_0$ the result is
\begin{equation}
\begin{aligned}
G_0^{ret}(\omega,\vec{r})=&f_1(\omega) H_0^{(1)}(\frac{r\sqrt{\omega^2-\Delta_0^2}}{\nu_F})\sgn{(\omega)}\\
&+f_2(\hat{r},\omega) H_1^{(1)}(\frac{r\sqrt{\omega^2-\Delta_0^2}}{\nu_F}),
\end{aligned}
\end{equation}
where $f_1(\omega)=-\frac{iv}{4\nu_F^2} [\omega-\Delta_0(\tau_x \cos{\phi}-\tau_y \sin{\phi}) ]$ and $f_2(\hat{r},\omega)=\frac{v}{4\nu_F^2 } (\vec{\sigma} \cdot \hat{r}) \tau_z\sqrt{\omega^2-\Delta_0^2}$, $H_{0,1}^{(1)}$ are the Hankel functions, $v$ is the volume of the lattice primitive cell, and $\hat{r}$ is the unit vector. For $\vec{r}$$\rightarrow$$0$, The Green's function takes the following asymptotic form:
\begin{equation}
G_0^{ret}(\omega,0)=f_1(\omega)\{\sgn{(\omega)}+ i\frac{2}{\pi}[\ln (\frac{\sqrt{\omega^2-\Delta_0^2}}{2W})+\chi]\},
\end{equation}
where $W$ is a large band cutoff, $\chi$ is the Euler-Mascheroni constant, and $\sgn{(x)}$ is the sign function. From the algebraic relations of Eqs.~(4) and (6), we can calculate the $T$-matrix, and further obtain the full retarded Green's function as:
\begin{equation}
G^{ret}(\omega,\vec{r})=G^{ret}_0(\omega,\vec{r})+G^{ret}_0(\omega,\vec{r}) \hat{T}(\omega)G^{ret}_0(\omega,-\vec{r}).
\end{equation}
The spin-resolved LDOS in direction $i$ is given as
\begin{equation}
\rho_{i}^{\pm}(\omega,\vec{r})=-\frac{1}{2\pi} \text{Im}\{\text{Tr}[G^{ret}(\omega,\vec{r})(1\pm \sigma_i)(1+\tau_z)]\}.
\end{equation}
\begin{figure}
\includegraphics[width=\columnwidth]{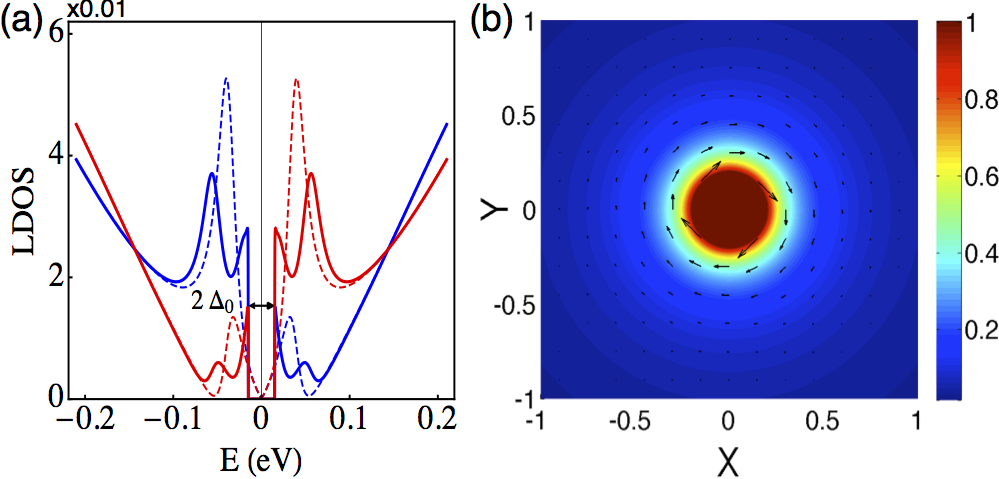}
\label{fig:local density of states}
\caption{(Color online) (a) LDOS as a function of the electron energy $E$ at the location $r=3$ nm away from a magnetic impurity. The solid and dashed lines are the spin-resolved LDOS for $\Delta_0=15$ meV and $0$ meV, respectively, with each of the spin-down (blue) and spin-up (red) resonance states split by the superconducting gap. (b) Spatial distribution of the spin-resolved LDOS at $E=0.2$ eV, with the arrow and color indicate the in-plane and $z$-direction projections, respectively. Here in order to highlight the resonance effect, we take a large value of $J=8.0$ eV \cite{R. R. Biswas}.}
\end{figure}

It was shown previously that a strong magnetic impurity will induce a pair of low-energy resonance states on the surface of a 3D TI \cite{R. R. Biswas}. In the present study, such LDOS resonances are derived from the minima of the denominators of the T-matrix. As illustrated in Fig.~1(a), these low-energy resonance states are further shown to be robust when the surface state of the 3D TI is proximity-coupled to the superconductor. Furthermore, each of the spin-resolved resonance states outside the superconducting gap will further be split due to the appearance of the superconducting gap around the Fermi level. Such resonance state splittings could be directly observed experimentally. The spatial distribution of the spin-resolved LDOS at a given energy is shown in Fig.~1(b). Similar to the case of TI surface states \cite{Q. Liu}, a magnetic impurity polarized along the vertical $z$ direction will induce spin textures both perpendicular and parallel to the interface due to the presence of SOC.

In order to study the emergent electronic properties inside the superconducting gap induced by the presence of the magnetic impurity, we calculate the retarded Green's function for $|\omega|<\Delta_0$:
$
G_0^{ret}(\omega,\vec{r})=\frac{2}{\pi}[-if_1(\omega) K_0^{(1)}(\frac{r\sqrt{\Delta_0^2-\omega^2}}{\nu_F})+f_2(\hat{r},\omega)K_1^{(1)}(\frac{r\sqrt{\Delta_0^2-\omega^2}}{\nu_F})],
$
where $K_{0,1}^{(1)}$ are the modified Bessel functions. For $\vec{r}$$\rightarrow$$0$, the above retarded Green's function takes the asymptotic form $G_0^{ret}(\omega,0)=i\frac{2}{\pi}f_1(\omega)c(\omega)$ with $c(\omega)=\ln{(\sqrt{\Delta_0^2-\omega^2}/2W)}+\chi$, which is a real function of $\omega$. From Eq.~(4), we find two poles of the $T$-matrix, giving rise to two impurity states inside the superconducting gap (not shown in Fig.~1(a)). For $J>0$, the impurity states can be obtained from the self-consistent relation: $\omega=\pm [\Delta_0 + \frac{\pi} {JSc(\omega)}]$. The symmetric nature of the two spin-up and spin-down impurity states stems from the particle-hole symmetry.

\textit{Multiple magnetic impurities}.---In this part, we focus on the electronic and magnetic properties of the superconductor-TI interface doped with randomly distributed magnetic impurities. In order to study the collective magnetic behavior of such a system, we first consider the RKKY interaction between two magnetic impurities mediated by the BCS qusi-particles. Hamiltonian (1) can be mapped into a two-band spinless $p_x+ip_y$ Hamiltonian as 
\begin{equation}
H_0=\sum_{km} \xi_{km} \alpha_{km}^{\dagger}  \alpha_{km}-\frac{1}{2} (m\Delta e^{i\theta_k}  \alpha_{km}^{\dagger}  \alpha_{-km}^{\dagger} +h.c.),
\end{equation}
where $\xi_{km}=m \nu_F k-\mu$ are the Dirac electron spectra, $m=\pm1$ are the band indices, and $\alpha_{km}=(me^{i\theta_k} c_{k \uparrow}+c_{k \downarrow})/\sqrt{2}$. Using the same basis set, Hamiltonian (2) can be rewritten as
\begin{equation}
H_{sd}^{i}=-J \sum_{mm'kk'} e^{i(\vec{k'}-\vec{k}) \cdot \vec{R}_i} ( \vec{S}_i \cdot \vec{\sigma}_{km;k'm'}) \alpha_{km}^{\dagger} \alpha_{k'm'},
\end{equation}
where $\vec{\sigma}_{km;k'm'}$ are the spin matrices. 

\begin{figure}
\includegraphics[width=\columnwidth]{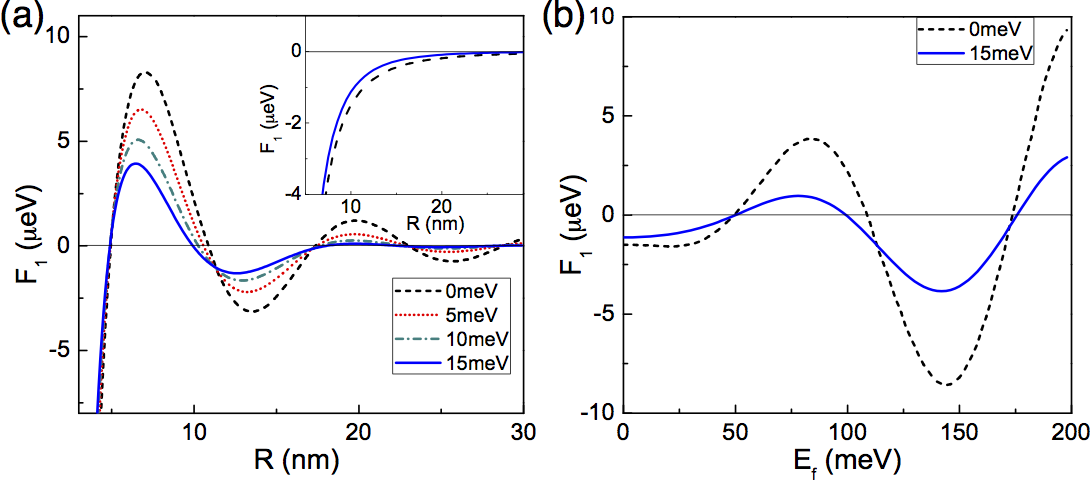}
\label{fig:local density of states}
\caption{(Color online) RKKY interaction between two magnetic impurities as a function of the separation $R$ and the Fermi energy $E_f$ calculated with $J=0.5$ eV, $E_f=100$ meV for (a) and $R=10$ nm for (b). The insert in (a) shows the case for the Fermi surface located at the Dirac point.}
\end{figure}

In the following, we treat the many-body problem using perturbation theory. The corrected ground state energy due to $s$-$d$ hybridization is $E=\langle \Omega|T H_0 S(-\infty,\infty)|\Omega \rangle$, where $T$ is the time-order operator, $|\Omega \rangle$ is the ground state of the BCS Hamiltonian (9), and the $S$-Matrix is defined as $S(t,t')=T \exp{[-i \int_{t'}^t dt_1 \hat{H}_{sd}(t_1)]}$. The normalized ground state of Hamiltonian (9) can be written as:
\begin{equation}
|\Omega \rangle=\prod_{km}{'}(u_{km}+\nu_{km} \alpha_{km}^{\dagger} \alpha_{-km}^{\dagger}) |0\rangle,
\end{equation}
where the sign $'$ indicates lack of double counting of electron pairs, $|0\rangle$ is the vacuum state, and $u_{km}$ and $\nu_{km}$ are determined by the Bogoliubov transformation. The normalization condition $\langle \Omega|\Omega \rangle=1$ is ensured by $|u_{km}|^2+|\nu_{km}|^2=1$. By expanding the $S$-matrix to the second order in $H_{sd}$, and only considering the loop approximation between two different magnetic impurities $i$ and $j$, the RKKY interaction can be effectively written as
\begin{equation}
\begin{aligned}
H_{ij}^{RKKY}=&F_1(R,\mu) \vec{S}_i \cdot \vec{S}_j +F_2(R,\mu) (\vec{S}_i \times \vec{S}_j)_x\\
&+F_3(R,\mu) S_i^x S_j^x,
\end{aligned}
\end{equation}
where 
\begin{equation}
\begin{aligned}
F_{\alpha}(R,\mu)=&-\frac{J^2v^2}{32\pi^2}  \sum_{mm'} \int_0^{k_c} dkdk' D_{km;k'm'}^{\alpha}(R) \\
  &\times \frac {kk'(E_{km}E_{k'm'}-\xi_{km}\xi_{k'm'}-\Delta_0^2)}{E_{km} E_{k'm'}(E_{km}+E_{k'm'})},
\end{aligned}
\end{equation}
with $\alpha=$1, 2, or 3, $k_c$ is a large momentum cutoff, $E_{km}=\sqrt{\xi_{km}^2+\Delta_0^2}$ is the excitation spectrum of the BCS quasi-particles, which can be obtained by diagonalizing Hamiltonian (9). In Eq.~(13), we also have $D_{km;k'm'}^{1}(R)=J_0(kR)J_0(k'R)-mm'J_1(kR)J_1(k'R)$, $D_{km;k'm'}^{2}(R)=m'J_0(kR)J_1(k'R)+mJ_1(kR)J_0(k'R)$, $D_{km;k'm'}^{3}(R)=2mm'J_1(kR)J_1(k'R)$, and $J_{0,1}(x)$ are the Bessel functions of the first kind. From Eq.~(12), we note that the RKKY interaction at the interface contains three different kinds: the Heisenberg-like term, the Dzyaloshinskii-Moriya (DM)-like term, and the Ising-like term. On a face level, the overall behavior is qualitatively similar to that on a TI surface \cite{J. J. Zhu} due to the SOC effects in both systems. However, the presence of the superconducting part introduces crucial differences, as reflected in Eq.~(13) and discussed in more detail below.

In general, the oscillation period of the RKKY interaction is determined by the Fermi wavelength $\lambda_F=1/k_F$. As shown in Fig.~2(a), an increase in the superconducting gap $\Delta_0$ suppresses both the oscillation magnitude and period of the RKKY interaction, exhibiting a fast decay of the long-rang part of the interaction to be close to zero. These behaviors stem from two physical aspects. First, the proximity-induced superconductivity will introduce a gap of $2\Delta_0$ at the Fermi level by forming Copper pairs; because every excitation of the quasi-particles has to overcome the superconducting gap, the corresponding RKKY interaction mediated by the quasi-particles will be suppressed in magnitude, especially the long-range part. Secondly, since the occupied states close to the superconducting gap dominate the contribution to the RKKY interaction, the corresponding wave vector is smaller than $k_F$, leading to a modification in the oscillation period. In Fig.~2(b), the Fermi energy dependence of the RKKY interaction is also presented.

\begin{table}
\caption{Robust Curie temperatures for systems of different superconducting gaps, obtained with x=0.03.}
\begin{tabularx}{\columnwidth}{p{0.1\columnwidth}<{\centering} p{0.2\columnwidth}<{\centering} p{0.2\columnwidth}<{\centering} p{0.2\columnwidth}<{\centering} p{0.2\columnwidth}<{\centering} }
\hline
\hline
$\Delta_0$ & 0meV & 5meV & 10meV & 15meV \\
$T_c^{MF}$ &  3.282K &  3.228K & 3.234K  & 3.243K \\
\hline
\hline
\end{tabularx}
\label{table:Curie temperature}
\end{table}

From the RKKY interaction described above, we can obtain the collective behavior of the magnetic impurities under the realistic assumption that their spatial distribution is random. The positional randomness combined with Eq.~(11) makes the in-plane interaction frustrated, while the ferromagnetic interactions between the $z$ components of the local spins can be optimized. Accordingly, a $z$-direction aligned ferromagnetic ground state is expected for the multiple magnetic impurity system, even though the atomic $s$-$d$ hybridization is isotropic. The mean-field virtual crystal approximation (MF-VCA) can be employed to estimate the Curie temperature $T_c^{MF}$, given as \cite{G. Bouzerar,H. Chen}
\begin{equation}
k_BT_c^{MF}=\frac{2x}{3} \sum_{i(i\neq0)} J_{0i},
\end{equation}
where the sum extends over the virtual sites, and $x$ is the concentration of the magnetic impurities on those virtual sites. The continuum limit is reached with $k_BT_c^{MF}=\frac{4\pi n_i}{3} \int_0^{\infty} rJ(r)dr$, where $n_i$ is the density of the magnetic impurities. For Bi$_2$Se$_3$, the virtual sites are the locations of the Bi atoms. By setting $x=3\%$, $a$=4.14 \AA$, J=0.5$ eV, and $E_f=0.1$ eV, the estimated $T_c$ for different superconducting gaps are listed in Table \ref{table:Curie temperature}. As shown in Fig.~2(a), the behaviors of the RKKY interaction are dramatically influenced by the superconducting gap, while the MF $T_c$ shows nearly constant values. These intriguing phenomena stem from a subtle compensation effect between ferromagnetism and anti-ferromagnetism: For $\Delta_0=0$, the magnitude of the long-range RKKY interaction shows a spatial dependence as $1/R^2$ \cite{J. J. Zhu}, favoring anti-ferromagnetism, while the short-range correlation always favors ferromagnetism. For $\Delta_0 \neq 0$, both the magnitude and long-range oscillation of the RKKY interaction will be suppressed, which again mutually compensate each other, leading to robust Curie temperatures as listed in Table \ref{table:Curie temperature}.

\textit{Chiral TSC and topological phase transition}.---Now we discuss the topological state of the superconductor-TI interface in the presence of random magnetic impurities. Based on the MF approximation, we first estimate the effective exchange field induced by the randomly distributed magnetic impurities, given by $V_{ex}=3Jx\langle S_z \rangle$. Within the picture that a given magnetic impurity interacts with an effective Zeeman field $B_{eff}=x\sum_iJ_{0i} \langle S_z \rangle$ defined by all the other magnetic impurities, its magnetic polarization is given by $\langle S_z \rangle=S B(\frac{B_{eff}S}{k_B T})$, where $B(x)=\frac{2S+1}{2S}\coth{(\frac{2S+1}{2S}x)}-\frac{1}{2S}\coth{(\frac{1}{2S}x)}$ is the Brillouin function. Therefore, the self-consistent solution of $\langle S_z \rangle$ and $B_{eff}$ can give rise to the temperature dependence of $V_{ex}$, as shown in Fig.~3(a). Importantly, the very existence of the $V_{ex}$ breaks the TRS by producing a gap at the Dirac point of the TI surface state, which in turn characterizes the chiral nature of the superconducting system, as further elaborated below. 

\begin{figure}
\includegraphics[width=\columnwidth]{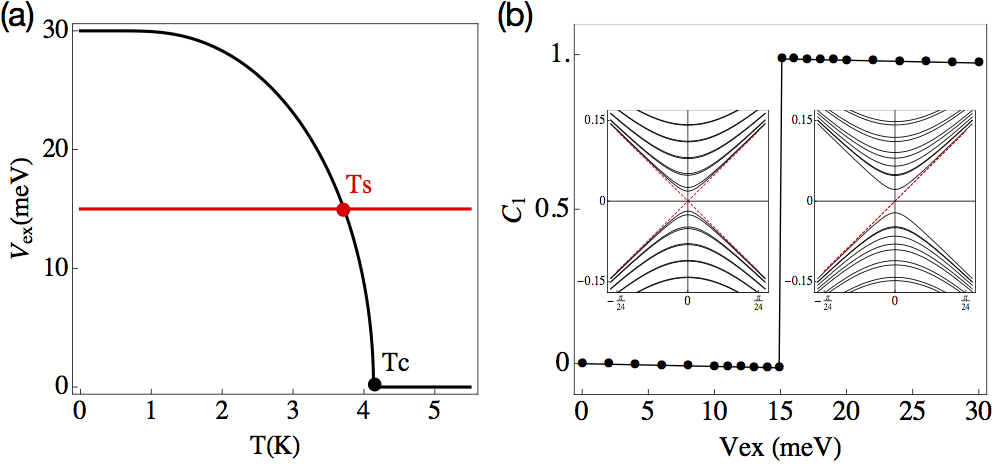}
\label{fig:local density of states}
\caption{(Color online) (a) The effective exchange field $V_{ex}$ (black) and superconducting gap $\Delta_0$ (red) as a function of the temperature. $T_s$ indicates the topological phase transition temperature, which is below the ferromagnetic Curie temperature $T_c$. (b) The first Chern number as a function of the effective exchange field $V_{ex}$. The insets illustrate the bulk band spectra (black solid) and edge states (red dashed) in the helical and chiral states, calculated with $V_{ex}=5$meV and $V_{ex}=25$meV, respectively. Other parameters include $\Delta_0=15$meV, $J=0.5$eV, $E_f=0$meV, and $x=0.03$.}
\end{figure}

In analogy with Ref.~\cite{L. Fu}, by defining the Bogliubov quasi-particle operators as $\gamma(\textbf{r})=\sum_{\sigma} u_{\sigma} (\textbf{r}) \psi^{\dagger}_{\sigma} (\textbf{r}) +\nu_{\sigma} (\textbf{r}) \psi_{\sigma} (\textbf{r})$, and solving the BdG equation $\mathcal{H}_{\text{BdG}} \Psi (\textbf{r})=E \Psi (\textbf{r})$ with $\Psi(\textbf{r})=[\nu_{\uparrow} (\textbf{r}) ,\nu_{\downarrow}(\textbf{r}),u_{\downarrow}(\textbf{r}),u_{\uparrow}(\textbf{r})]^T$ at geometrical edges, we can find two types of Majorana edge states by varying $V_{ex}$. First, for $\sqrt{\mu^2+\Delta_0^2}>V_{ex}>\mu$, there are two helical edge states, given by $\Psi_{\pm} (x)=\frac{1}{\mathcal{N}_{\pm}} (\sqrt{V^{-}},\mp i \sqrt{V^{+}},\mp e^{i\phi}  \sqrt{V^{+}},ie^{i\phi}  \sqrt{V^{-}} )^T e^{-\eta x}$, where $V^{\pm}=V_{ex}\pm \mu$, $\eta=(\Delta_0 \pm \sqrt{V^{+}V^{-}})/\nu_F$ and $\mathcal{N}_{\pm}$ are the normalization parameters. It is easy to verify $\gamma^{\dagger}(k_y)=\gamma(-k_y)$, which implies that the solutions are Majorana edge modes. In order to give an intuitional picture of the helical edge states, we evaluate the low-energy ``$k \cdot p$" Hamiltonian as $\mathcal{H}_h=\sqrt{1-(\mu/V_z)^2} \nu_F k_y \tau_z$, where $\tau_z$ is the Pauli matrix. Secondly, for $V_{ex}>\sqrt{\mu^2+\Delta_0^2}$, there are two degenerate chiral Majorana edge states $\Psi_{\pm} (x)=\frac{1}{\mathcal{N}_{\pm}} (\sqrt{V^{-}},-i \sqrt{V^{+}},\mp e^{i\phi}  \sqrt{V^{+}},\pm ie^{i\phi}  \sqrt{V^{-}} )^T e^{\mp \eta x}$, where $\mathcal{N}_{\pm}$ are the normalization parameters. The chiral nature can be illustrated by the low-energy ``$k\cdot p$" Hamiltonian, given by $\mathcal{H}_c=\sqrt{1-(\mu/V_z)^2} \nu_F k_y $. Therefore, by varying the exchange field $V_{ex}$, we can expect a topological phase transition from a helical to chiral TSC at $V_{ex}=\sqrt{\mu^2+\Delta_0^2}$, and the corresponding transition temperature is marked by $T_s$ in Fig.~3(a).

As a quantitative measure for the occurrence of the topological phase transition, we calculate the first Chern number for systems before and after the transition. The first Chern number can be defined as the integral of the Berry curvature over the first Brillouin Zone \cite{D. Xiao}: $\mathcal{C}_1=\frac{1}{2\pi} \int_{BZ} (\partial_{k_x} \mathcal{A}_{k_y}-\partial_{k_y} \mathcal{A}_{k_x}) \text{d} \textbf{k}$, where $\mathcal{A}_{k_{\alpha}}=-i \sum_n \langle u_n(\textbf{k}) |\partial_{k_{\alpha}}| u_n(\textbf{k})\rangle$ is the Berry connection, $\alpha=x,y$, and the index $n$ runs over all the occupied states. Hamiltonian (1) can be regularized on a square lattice with the substitution $p_{x,y}\rightarrow a^{-1} \sin{(p_{x,y}a)}$, where $a$ is the lattice constant. The results for $\mu=0$ are shown in Fig.~3(b).  There are two sets of subbands due to spin degrees of freedom. When $V_{ex}<\Delta_0$, the resulting Chern numbers from the two sets are equal in magnitude but opposite in sign, and the total Chern number $\mathcal{C}_1=0$ signifies a non-chiral TSC state. When $V_{ex}>\Delta_0$, one set of the subbands will be inverted by the exchange field, and the corresponding Chern number will also reverse sign, resulting in $\mathcal{C}_1=1$, indicating a chiral TSC state. 

So far, we have focused on realizing chiral TSC at the superconductor-TI interface. As a natural  extension, here we also briefly discuss the proposed scheme in connection with recent experiments \cite{M. X. Wang,E. Wang}.  In particular, when Bi$_2$Se$_3$ was grown on the $d$-wave superconductor of Bi$_2$Sr$_2$CaCu$_2$O$_{8+\delta}$, an $s$-wave superconducting gap as large as 15 meV was observed on the top surface of the TI \cite{E. Wang}. Based on these experiments, we expect that the proposed mechanism can also be exploited to realize chiral TSCs on tops of TI/superconductor heterostructures. 

In summary, we have proposed an alternative and conceptually simpler scheme to realize a chiral TSC, achieved by doping magnetic impurities directly at a superconductor-TI interface. 
We have found that, for randomly distributed magnetic impurities, the RKKY interaction gives rise to a persistent  ferromagnetic state independent of the superconducting gap. The ferromagnetic state can naturally provide a uniform and strong exchange field, which in turn breaks the time reversal symmetry, driving the system from a helical TSC phase into a chiral TSC phase at sufficiently low temperatures. The proposed scheme is in principle also applicable        on top of a TI/superconductor heterostructure, or when the TI is replaced by a normal semiconductor with strong Rashba SOC. These findings therefore provide new platforms for realizing chiral TSC, observing Majorana zero modes, and executing TQC.

This work was supported by the NSFC Grant No.11034006 and National Key Basic Research Program of China (2014CB921103).


\begin{thebibliography}{99}
\bibitem{A. Kitaev}A. Y. Kitaev, Ann. Phys. (N.Y.) \textbf{303}, 2 (2003).
\bibitem{C. Nayak}C. Nayak, S. H. Simon, A. Stern, M. Freedman, S. Das Sarma, Rev. Mod. Phys. \textbf{80}, 1083 (2008).
\bibitem{G. Moore}G. Moore and N. Read, Nucl. Phys. \textbf{B360}, 362 (1991).
\bibitem{X. L. Qi1}X. L. Qi and S. C. Zhang, Rev. Mod. Phys. \textbf{83}, 1057 (2011).
\bibitem{L. J. Buchholtz}L. J. Buchholtz and G. Zwicknagl, Phys. Rev. B \textbf{23}, 5788 (1981).
\bibitem{M. Matsumoto}M. Matsumoto and M. Sigrist, J. Phys. Soc. Jpn. \textbf{68}, 994 (1999).
\bibitem{X. L. Qi2}X. L. Qi, T. L. Hughes, and S. C. Zhang, Phys. Rev. B \textbf{82}, 184516 (2010).
\bibitem{D. A. Ivanove}D. A. Ivanov, Phys. Rev. Lett. \textbf{86}, 268 (2001).
\bibitem{N. Read}N. Read and D. Green, Phys. Rev. B \textbf{61}, 10267 (2000).
\bibitem{S. Das Sarma}S. Das Sarma, C. Nayak, and S. Tewari, Phys. Rev. B \textbf{73}, 220502(R) (2006).
\bibitem{L. Fu}L. Fu and C. L. Kane, Phys. Rev. Lett. \textbf{100}, 096407 (2008); Phys. Rev. B \textbf{79}, 161408 (2009).
\bibitem{J. D. Sau}J. D. Sau, R. M. Lutchyn, S. Tewari, and S. Das Sarma, Phys. Rev. Lett. \textbf{104}, 040502 (2010).
\bibitem{J. Alicea}J. Alicea, Phys. Rev. B \textbf{81}, 125318 (2010).
\bibitem{V. Mourik}V. Mourik \textit{et al.}, Science \textbf{336}, 1003 (2012).
\bibitem{A. Das}A. Das \textit{et al.}, Nature Phys. \textbf{8}, 887 (2012).
\bibitem{M. T. Deng}M. T. Deng \textit{et al.}, Nano Lett. \textbf{12}, 6414 (2012).
\bibitem{D. Xiao}D. Xiao, M. C. Chang, and Q. Niu, Rev. Mod. Phys. \textbf{82}, 1959 (2010).
\bibitem{H. J. Zhang} H. J. Zhang \textit{et al.}, Nature Phys. \textbf{5}, 438 (2009).
\bibitem{Q. Liu}Q. Liu, C. X. Liu, C. K. Xu, X. L. Qi, and S. C. Zhang, Phys. Rev. Lett. \textbf{102}, 156603 (2009).
\bibitem{J. S. Dyck}J. S. Dyck, P. Hajek, P. Lostak, and C. Uher, Phys. Rev. B \textbf{65}, 115212 (2002).
\bibitem{J. J. Zhu}J. J. Zhu, D. X. Yao, S. C. Zhang, and K. Chang, Phys. Rev. Lett. \textbf{106}, 097201 (2011).
\bibitem{A. V. Balatsky}A. V. Balatsky, I. Vekhter, and J. X. Zhu, Rev. Mod. Phys. \textbf{78}, 373 (2006).
\bibitem{R. R. Biswas}R. R. Biswas and A. V. Balatsky,  Phys. Rev. B \textbf{81}, 233405 (2010).
\bibitem{G. Bouzerar}G. Bouzerar, J. Kudrnovsky, L. Bergqvist, and P. Bruno, Phys. Rev. B \textbf{68}, 081203(R) (2003).
\bibitem{H. Chen}H. Chen, W. G. Zhu, E. Kaxiras, and Z. Y. Zhang, Phys. Rev. B \textbf{79}, 235202 (2009).
\bibitem{M. X. Wang}M. X. Wang \textit{et al.}, Science \textbf{336}, 52 (2012).
\bibitem{E. Wang}E. Wang \textit{et al.}, Nature Phys. \textbf{9}, 621 (2013).
\end{thebibliography}
\end{document}